\documentclass[printer]{aa} 
\usepackage{graphicx}
\usepackage{natbib}  

\usepackage{txfonts}
\begin{document}
\authorrunning{Petrovich \and Reisenegger}
\titlerunning{Long-period thermal oscillations in superfluid MSPs.}
     \title{Long-period thermal oscillations in superfluid millisecond pulsars.}

   \author{Cristobal Petrovich\footnotemark[1]
                  \and
          Andreas Reisenegger }

   \institute{Departamento de Astronom\'\i a y Astrof\'\i sica, 
		Pontificia Universidad Cat\'olica de Chile,
		Casilla 306, Santiago 22, Chile.}



  \abstract
   {In previous papers, we have shown that, as the rotation of a neutron star slows
down, it will be internally heated as a consequence of the progressively changing mix
of particles (rotochemical heating). In previously studied cases (non-superfluid neutron
stars or superfluid stars with only modified Urca reactions), this leads to a quasi-steady
state in which the star radiates thermal photons for a long time, possibly accounting
for the ultraviolet radiation observed from the millisecond pulsar J0437-4715.
}
   {   For the first time, we explore the phenomenology of rotochemical heating with
direct Urca reactions and uniform  and isotropic superfluid energy gaps of different sizes.
    }
   {We first do exploratory work by integrating the thermal and chemical evolution
equations numerically for different energy gaps, which uncovers a rich phenomenology
of stable and unstable solutions. To understand these, we perform a stability analysis
around the quasi-steady state, identifying the characteristic times of growing, decaying,
and oscillating solutions.}
  { For small gaps, the phenomenology is similar to the previously studied cases,
in the sense that the solutions quickly converge to a quasi-steady state. For large gaps
($\gtrsim 0.05$ MeV), these solutions become unstable, leading to a limit-cycle behavior
of periodicity $\sim 10^{6-7}$ yr, in which the star is hot ($T_s\gtrsim 10^5$ K) for
a small fraction of the cycle ($\sim 5- 20 \%$ ), and cold for a longer time.}
   {}

\keywords{stars: neutron --- dense matter --- stars: rotation
--- pulsars: general --- pulsars: individual (PSR J0437-4715)}

   \maketitle
%

\section{Introduction} \label{sec:intro}

The observation of thermal emission from the surface of a neutron star (NS) has the potential
to provide constraints on its inner structure. In the existing literature, several detailed cooling 
calculations have been compared to the few estimates available for the surface temperatures 
of neutron stars (see \citealt{yak04} for a review and references). These calculations 
are based on passive cooling, at first neutrino-dominated, and later driven by photon emission at ages $\gtrsim10^5$ yr.

Several proposed mechanisms could keep NSs hot beyond the standard
cooling timescale $\sim10^7$ yr
(see, e.g., \citet{Schaab}, but note that this paper gives an incorrect parameterization
of rotochemical heating, as explained in \citealt{denis10}).  
 \citet{denis10} compared the abilities of these mechanisms to 
reheat NSs, concluding that
two of them might be  important  in old NSs, namely, reheating by frictional motion of superfluid neutron vortices (\citealt{alpar}; \citealt{shibazaki}) and
\textit{rotochemical heating}.
The latter was first proposed
by \citet{reis95} and then improved by \citet{FR05}, and considered
the internal structure of non-superfluid NSs with realistic equations of state
(EOSs) in the framework of general relativity.
It works as follows. As the NS's rotation rate decreases, the
reduction in the centrifugal force makes it contract. 
This perturbs each fluid element, raising the local pressure and causing deviations 
from beta equilibrium. Eventually, the system reaches a quasi-steady configuration, where 
the rate at which spin-down modifies the equilibrium concentrations is the same as that
at which neutrino reactions restore the equilibrium. These reactions heat the stellar
interior, making the star emit thermal radiation.
\footnotetext[1]{\textit{Present address}: Department of Astrophysical Sciences, Princeton University, Princeton, NJ 08544, USA,
 \email{cpetrovi@astro.princeton.edu}}
The potential presence of superfluid nucleons in the NS's interior 
has been widely considered to model their thermal evolution
(see, e.g. \citealt{yak04} for cooling models)
since it considerably reduces the neutrino reactions and the specific heat 
involving superfluid species \citep{yak01},  
and opens new neutrino emission processes, 
namely pair breaking and formation reactions \citep{flowers}. 

In a previous paper \citep{petro10}, we modeled rotochemical heating 
of millisecond pulsars with only modified Urca reactions in the 
presence of uniform and isotropic Cooper pairing gaps of neutrons $\Delta_{n}$ and protons
$\Delta_{p}$.
We verified the order-of-magnitude predictions of \citet{reis97}, 
 finding that the chemical imbalances in the star grow up to the 
threshold value $\Delta_{thr}=\mbox{min}(  \Delta_n+3\Delta_p,3\Delta_n+\Delta_p )$,
 which is higher than in the quasi-steady state achieved in the absence of superfluidity. 
 Therefore, the old superfluid NSs will take longer to reach the quasi-steady
state than their nonsuperfluid counterparts, and they have a higher 
a luminosity in this state, given
by $L_\gamma^{\infty,qs} \simeq \left(1-4\right) \times 10^{32}\left(\Delta_{thr}/\mbox{MeV}\right)
\left(\dot{P}_{-20}/P_{\mbox{\small{ms}}}^3\right)\mbox{ erg}\mbox{ s}^{-1}$, where
$\dot{P}_{-20}$ is the period derivative in units
of $10^{-20}$ and $P_{\mbox{\small{ms}}}$ is the period in milliseconds.
With the previous relation, we  found that energy gaps in the range
$ 0.05 [\mbox{MeV}]\lesssim\Delta_{thr}\lesssim0.45 [\mbox{MeV}]$
are consistent with the ultraviolet emission of PSR J0437-4715
\citep{kargaltsev04}.

We extend our previous analysis to the case
where the much faster direct Urca reactions are allowed.
We find a qualitatively new behavior of rotochemical
heating, where the temperature and chemical imbalances \textit{oscillate}
around the quasi-steady state.

The structure of this paper is the following. 
In Sect. \ref{sec:theory}, we present the basic
equations of rotochemical heating and the 
phase-space integrals for direct Urca reactions with Cooper pairing
gaps and chemical imbalances.
In Sect. \ref{sec:results}, we describe our 
 results for the thermal evolution and the linear
stability analysis of old NSs.
We summarize our main conclusions in Sect. \ref{sec:conclusions}.

		\section{THEORETICAL FRAMEWORK}\label{sec:theory}

\subsection{Rotochemical heating: basic equations} \label{sec:theory_basic}

The basic framework for rotochemical heating is explained in detail
in \citet{FR05}, and the modifications made to consider the Cooper pairing effects are
described in \citet{petro10}. The latter is the framework we use throughout  this paper.
Therefore, we just point out the fundamental equations for 
completeness and to clarify the notation of the present paper.
 
We consider the simplest model of a neutron star core, composed
of neutrons, protons, electrons, and muons ($npe\mu$ matter), ignoring
the potential presence of exotic particles.

The internal temperature, redshifted to a distant observer, 
$T_\infty$, is taken to be uniform inside the star because
we are modeling the thermal evolution over timescales
much longer than the diffusion time \citep{reis95}. Thus, the evolution of 
the internal temperature for an isothermal interior is given by the thermal 
balance equation \citep{thorne77} 
\begin{eqnarray}
\label{eq:evolucion_Ti}
\dot{T}_\infty & =& \frac{1}{C} \left( L_H^{\infty}-L_\nu^\infty - L_\gamma^\infty\right),
\end{eqnarray}
where $C$ is the total heat capacity of the star, $L_H^{\infty}$ is the
total power released by the heating mechanism, $L_\nu^\infty$ the total 
power emitted as neutrinos due to Urca reactions,
and $L_\gamma^\infty$ the power released as thermal photons.

The amount of energy released by each Urca-type reaction is 
$\eta_{npl}=\mu_n-\mu_p- \mu_l$ ( $l=e,\mu$), where $\mu_{i}$ is the
chemical potential of the particle species $i$.
Thus, we write the total energy dissipation rate as
\begin{eqnarray}
\label{eq:heating_term}
L_H^{\infty}= \eta_{npe}^{\infty}\Delta\tilde{\Gamma}_{npe}+\eta_{np\mu}^{\infty}\Delta\tilde{\Gamma}_{np\mu},
\end{eqnarray}
where $\Delta\tilde{\Gamma}_{npl}=\tilde{\Gamma}_{n\rightarrow pl}-\tilde{\Gamma}_{pl\rightarrow n}$ 
is the net reaction rate of the Urca reaction integrated over the core 
 involving the lepton $l$.

The photon luminosity is calculated 
by assuming black-body radiation $L_\gamma^\infty=4\pi\sigma R_\infty^2 T_{s,\infty}^4$,
where  $R_\infty$ and $T_{s,\infty}$ are the radius and the surface temperature 
of the star measured by an observer at infinity, respectively. 
To relate the internal and the surface temperatures,
the fully accreted envelope model of \citet{potetal97} is used.

The evolution of the redshifted chemical imbalances, also uniform throughout the core, 
is given by
\begin{eqnarray}
\label{eq:evolucion_eta1}
\dot{\eta}^\infty_{npe} &=&  -Z_{npe}\Delta\tilde{\Gamma}_{npe}
- Z_{np}\Delta\tilde{\Gamma}_{np\mu} + 2W_{npe}\Omega \dot{\Omega},\\
\label{eq:evolucion_eta2}
\dot{\eta}^\infty_{np\mu}  &=&  -Z_{np}\Delta\tilde{\Gamma}_{npe}
- Z_{np\mu}\Delta\tilde{\Gamma}_{np\mu} + 2W_{np\mu}\Omega \dot{\Omega},
\end{eqnarray}
where the terms $Z_{np}$, $Z_{npe}$, $Z_{np\mu}$, $W_{npe}$, and $W_{np\mu}$
are constants that depend on the stellar structure and are kept unchanged 
with respect to their latest definition in \citet{reis06}, and $\Omega \dot{\Omega}$
is the product of the angular velocity and its time
derivative (proportional to the spin-down power). 

As \citet{petro10} showed, the results of the evolution with rotochemical heating when computing  
$L_\nu^\infty$ and $\Delta\tilde{\Gamma}_{npe}$ in the presence of superfluid nucleons
are substantially different from their superfluid counterparts calculated by \citet{FR05}, 
since superfluidity strongly inhibits these reactions. Thus, the chemical imbalances become larger during the quasi-steady state, lengthening the timescale to arrive at
this state 
compared with the non-superfluid case, and predicting higher temperatures in old NSs.
In this paper, we include the powerful direct Urca reactions to our previous study.

\subsection{Cooper pairing}
\label{sec:theory_cooper}

In the core, neutrons are believed to form Cooper pairs because of their interaction in the 
triplet $^3P_2$ states via the anisotropic channels $|m_J|=0$ (type B) or $|m_J|=2$ (type C), while protons form  isotropic, singlet $^1S_0$ pairs (type A)
\citep{yak01}. 
Additionally, in the outermost core and inner crust, neutrons are believed to form 
singlet-state $^1S_0$ pairs. 
The $^3P_2$ (type B and C) state description is rather uncertain in the sense 
that the energetically most probable state of $nn$-pairs ($|m_J|=0,1,2$) is not known, 
being extremely sensitive to the still unknown $nn$-interaction
(see, e.g. \citealt{amundsen}).
Taking this classification into account, \citet{villain} solve numerically the 
suppression due to each type of superfluidity of the net reaction rate for 
direct Urca and modified Urca reactions out of beta 
equilibrium, finding that 
the suppression due to type A superfluidity is of strength between 
the suppression due to anisotropic channels  type B and type C
superfluidity, respectively.
For simplicity, we consider the energy gaps 
for the neutrons $\Delta_n$ and the protons $\Delta_p$ at zero temperature,
 redshifted to a distant observer, as parameters that are 
isotropic ($^1S_0$ pairs) and uniform throughout the core of the NS.

The phase transition for a nucleon species into a superfluid state takes place when
its  temperature falls below a critical value $T_c$. 
This temperature is related to the energy gap at zero temperature $\Delta(T=0)$; 
for the isotropic pairing channel $^1S_0$, 
$\Delta(T=0)=1.764 k T_c$. Additionally, when the transition occurs,
the amplitude of the energy gap depends on the temperature by means of the 
BCS equation \citep{yak01}, which can be fitted by the practical formula of
\citet{levyak94} for the isotropic gap
\begin{eqnarray}
\label{eq:delta}
 \delta \equiv \frac{\Delta(T)}{kT}=\sqrt{1-T/T_c}\left(1.456-\frac{0.157}{\sqrt{T/T_c}}+\frac{1.764}{T/T_c}\right),
\end{eqnarray}
where $\delta$ is the variable used in the phase-space integrals in
Sect. \ref{sec:theory_emiss}.
It is straightforward to check that the limiting cases are reproduced by Eq.
(\ref{eq:delta}), i.e. $\delta=0$ when $T=T_c$ and $\delta=\Delta(T=0)/kT$ when $T\ll T_c.$. 
 \citet{levyak94} claim that intermediate values of $T/T_c$ are also
reproduced by this formula with a maximum error less than $5\%$, which is accurate
enough for the purposes of this work.

Having defined the energy gap of the nucleon $\Delta_i$ with $i=n,p$, it is possible 
to express the momentum dependence of the nucleon energy $\epsilon_i(p_i)$ near the Fermi
level, i.e. $|p_i-p_{F_i}|\ll p_{F_i}$,  as follows \citep{yak01}
\begin{eqnarray}
\label{eq:dispertion}
 \epsilon_i(p_i)=\mu_i- \sqrt{v_{F_i}^2(p_i-p_{F_i})^2+\Delta_i^2}\quad\mbox{if}\quad p_i<p_{F_i} ,\nonumber\\ 
 \epsilon_i(p_i)=\mu_i+ \sqrt{v_{F_i}^2(p_i-p_{F_i})^2+\Delta_i^2}\quad\mbox{if}\quad p_i>p_{F_i},
\end{eqnarray}
where $p_{i}$, $p_{F_i}$, $v_{F_i}$, and  $\mu_i$ are the momentum, 
the Fermi momentum, the Fermi velocity,
and the chemical potential of species $i=n,p$, respectively.

\subsection{Neutrino emissivity}
\label{sec:theory_emiss}

The fastest reactions in NS cores are the direct 
Urca processes 
\begin{eqnarray}
n\rightarrow p+e^-+\bar{\nu}_l \\
p+e^-\rightarrow n+\nu_l,
\end{eqnarray}
where $l=e, \mu$. 
If these reactions are kinematically allowed, their emissivity is
much greater than those produced by the
modified Urca reactions \citep{yak01}.

We write the neutrino emissivity and the net 
reaction rate of direct Urca reactions involving the lepton $l$ and 
 integrated over the core, respectively, as
\begin{eqnarray}
\label{eq:L_noneq}
L_{\nu,l}^\infty & = & \tilde{L}_{l}I_{D,\epsilon} T_\infty^6,\\
\label{eq:gamma_noneq}
\Delta\tilde{\Gamma}_{npl} & = & \frac{\tilde{L}_{l}}{k}I_{D,\Gamma}T_\infty^5,
\end{eqnarray}
where constants $\tilde{L}_{nl}$ and $\tilde{L}_{pl}$ are defined in terms 
of the neutrino luminosities for a nonsuperfluid NS in beta equilibrium, as
\begin{eqnarray}
\label{eq:int_Qeq}
\tilde{L}_\alpha & \equiv & \frac{L_\alpha^{eq}}{T_\infty^6} = \int_{\mbox{core}} 4\pi r^2 e^{\Lambda} S_\alpha(n) e^{-4\Phi} dr,
\end{eqnarray}
where the term $S_l$ is a slowly varying function of the 
baryon number  density $n$ (e.g., \citealt{yak01}), and $\Lambda$ and $\Phi$ are the usual 
Schwarzschild metric terms. 
The quantities $I_{D,\epsilon}$ and $I_{D,\Gamma}$	
are dimensionless phase-space integrals 
that contain the dependence of the emissivity and the net reaction rate, 
respectively, on the chemical imbalances $\eta_{npl}^\infty$ and the 
energy gaps $\Delta_n$ and $\Delta_p$.

To introduce these integrals, it is useful to define the
usual dimensionless variables normalized by the thermal energy $kT$, as follows
\begin{equation}
\label{eq:adim_var_1}
 x_j\equiv \frac{\epsilon_j-\mu_j}{kT}  \mbox{ , }\mbox{  } x_\nu\equiv \frac{\epsilon_\nu}{kT},
\mbox{  }\mbox{ and } \mbox{  } \xi_l\equiv \frac{\eta_{npl}}{kT}  ,
\end{equation}
which represent the energy of the non-superfluid degenerate particle $j$, 
the neutrino, and the chemical imbalance involving
the lepton $l$, respectively, while for the superfluid nucleon $i$ we write
\begin{equation}
\label{eq:adim_var_2}
 x_i\equiv\frac{v_{F_i}(p_i-p_{F_i})}{kT}  \mbox{ and } 
z_i \equiv \mbox{sgn}(x_i)\sqrt{x_i^2+\delta_i^2},
\end{equation}
where $\delta_i$ is defined in Eq. (\ref{eq:delta}) in terms of $\Delta_{i}$.
Thus, in terms of these variables,
\begin{eqnarray}
\label{eq:integral_durca_emis}
I_{D,\epsilon}=\frac{914}{5040\pi^6}&&\int_0^{\infty}dx_{\nu} x_{\nu}^3
 \int_{-\infty}^{\infty} \int_{-\infty}^{\infty}\int_{-\infty}^{\infty}
dx_{n}dx_{p}dx_{e}\nonumber \\
&\times&f(z_n) f(z_p)  f(x_e) \left[\delta(x_{\nu}+\xi_l-z_n-z_p-x_e)  \right. \nonumber \\
&+&\left. \delta(x_{\nu}-\xi_l-z_n-z_p-x_e) \right] 
\end{eqnarray}
and
\begin{eqnarray}
\label{eq:integral_durca_gamma}
I_{D,\Gamma}=\frac{914}{5040\pi^6}&&\int_0^{\infty}dx_{\nu} x_{\nu}^2
 \int_{-\infty}^{\infty} \int_{-\infty}^{\infty}\int_{-\infty}^{\infty}
dx_{n}dx_{p}dx_{e}\nonumber \\
&\times&f(z_n) f(z_p)  f(x_e) \left[\delta(x_{\nu}+\xi_l-z_n-z_p-x_e)  \right. \nonumber \\
&-&\left. \delta(x_{\nu}-\xi_l-z_n-z_p-x_e) \right],
\end{eqnarray}
where  $f(\cdot)$ is the Fermi function $f(x)=1/(1+e^x)$, and the numerical
factor in front of the integral is normalizes $I_{D,\epsilon}$ to 1  when the
energy gaps and the chemical imbalances are zero.

In the nonsuperfluid case (i.e. $\delta_n=\delta_p=0$), 
these integrals reduce to 
the polynomials calculated by \citet{reis95}
\begin{eqnarray}
\label{eq:F_D_def}
I_{D,\epsilon} (\delta_n=\delta_p=0)=F_D(\xi_l)&=& 
 1 + \frac{1071\xi_l^2}{457\pi^2} + \frac{315\xi_l^4}{457\pi^4} + \frac{21\xi_l^6}{457\pi^6},\\
\label{eq:H_D_def}
I_{D,\Gamma} (\delta_n=\delta_p=0)=H_D(\xi_l)&=& 
\frac{714\xi_l}{457\pi^2}+\frac{420\xi_l^3}{457\pi^4}+\frac{42\xi_l^5}{457\pi^6}.
\end{eqnarray}

The phase-space integrals $I_{D,\epsilon}$ and $I_{D,\Gamma}$ cannot
be solved analytically when one or two nucleon species are superfluid.
Thus, \citet{villain} and \citet{petro10} computed these integrals numerically and described 
the suppression produced by the Cooper pairs by means of the so-called reductions factors.
The latter are defined as the ratio of the 
phase-space integrals with superfluid particle species to the non-superfluid
integrals, i.e. $R_{D,\epsilon}=I_{D,\epsilon} /F_D(\xi_l)$ 
and $R_{D,\Gamma}=I_{D,\Gamma} /H_D(\xi_l)$.
Hereafter, we calculate these integrals numerically by using the numerical method 
explained in \citet{petro10}, which is based on the Gauss-Laguerre quadrature,
taking advantage of the exponentially decaying behavior of the Fermi functions 
appearing in the integrand.

\section{RESULTS AND DISCUSSION} \label{sec:results}

\subsection{Evolution}\label{sec:results_evol}

In this section, we show the peculiar behavior of rotochemical heating in MSPs with
superfluid nucleons and direct Urca reactions. To do so, we numerically
compute the evolution of Eqs.
(\ref{eq:evolucion_Ti}), (\ref{eq:evolucion_eta1}), and (\ref{eq:evolucion_eta2}).
Hereafter, we omit  the $\infty$ subscript in the temperature and 
the $\infty$ superscript in the chemical imbalances to simplify the notation.

The NS interior structure is modeled by the BPAL21 EOS 
of \citet{pal88}   for the core,  
supplemented with those of \citet{prl95} and \citet{hanpichon94} for the inner and outer
crust, respectively. 
This model opens electron direct Urca reactions in the core when the NS mass is
greater than 1.67 $M_\odot$ and muon direct Urca reactions when its mass is
greater than 1.69  $M_\odot$,  and reaches a maximum mass value of 
1.71  M$_\odot$.
Thus, the direct Urca reactions are allowed within
 the mass range $1.76\pm0.2$ M$_\odot$  
of the MSP J0437-4715  \citep{verbiest}, the only MSP whose thermal 
radiation has been measured so far and, therefore, there is an estimate
of its surface temperature \citep{kargaltsev04}. 

For numerical calculations, the evolution of $\Omega\dot{\Omega}$
is computed by assuming magnetic dipole braking with 
no field decay, and relating the magnetic field on the magnetic equator, rotation period, and period derivative by the conventional formula $B=3.2 \cdot10^{19}(P\dot{P})^{1/2}$ G, where $P$ is measured in seconds.
The magnetic field is chosen to be $B=2.8\cdot10^8$ G, so as to match the currently
measured values of $P$ and $\dot{P}$ of PSR J0437-4715.

In Fig. 1, we show the evolution for three different values of the neutron energy gap,
with non-superfluid protons and no muons, 
illustrating the appearance of strong thermal oscillations at larger gaps.
The left top panel, where $\Delta_n=0.01$ MeV,  shows that 
rotochemical heating converges rapidly to a stable solution given by the
quasi-steady solution calculated from $\dot{T}=\dot{\eta}_{npe}=0$,
 as it happens when no superfluid 
nucleon species are considered \citep{FR05} or when superfluid
nucleons are present, but only modified 
Urca reactions are allowed \citep{petro10}.
A hint of an oscillation is seen before the solution converges.

In the left bottom panel of Fig. 1, we increase 
the value of the energy gap to $\Delta_n=0.03$ MeV.
In this case, there is a clearly visible damped oscillation
before the variables settle to their quasi-state values.

\begin{figure*}[!t]
\centering
\begin{tabular}{cc}
\includegraphics[width=8.15cm,height=5cm]{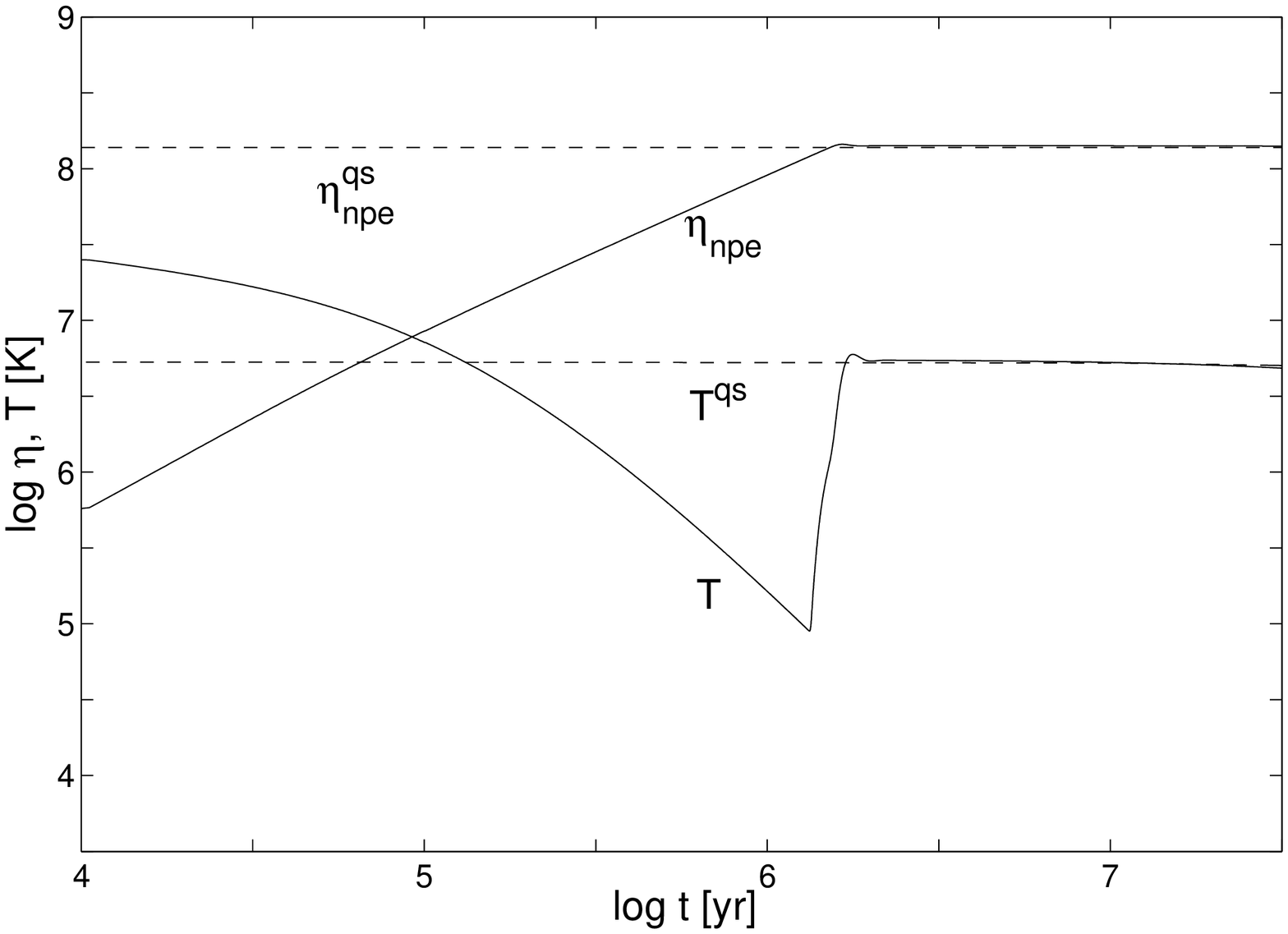}
\includegraphics[width=8.2cm,height=5cm]{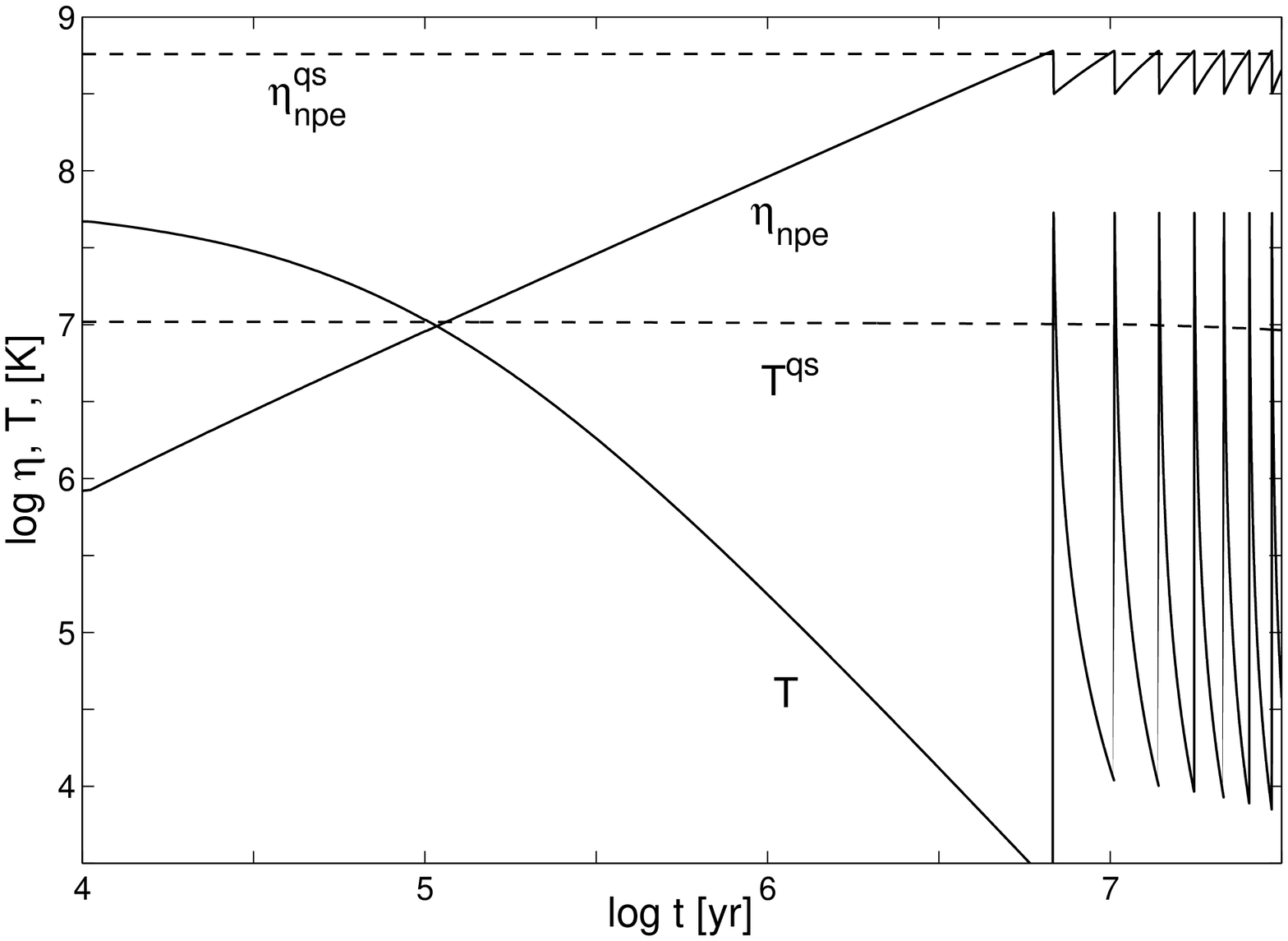} 
\end{tabular}
\begin{tabular}{cc}
\includegraphics[width=8.2cm,height=5cm]{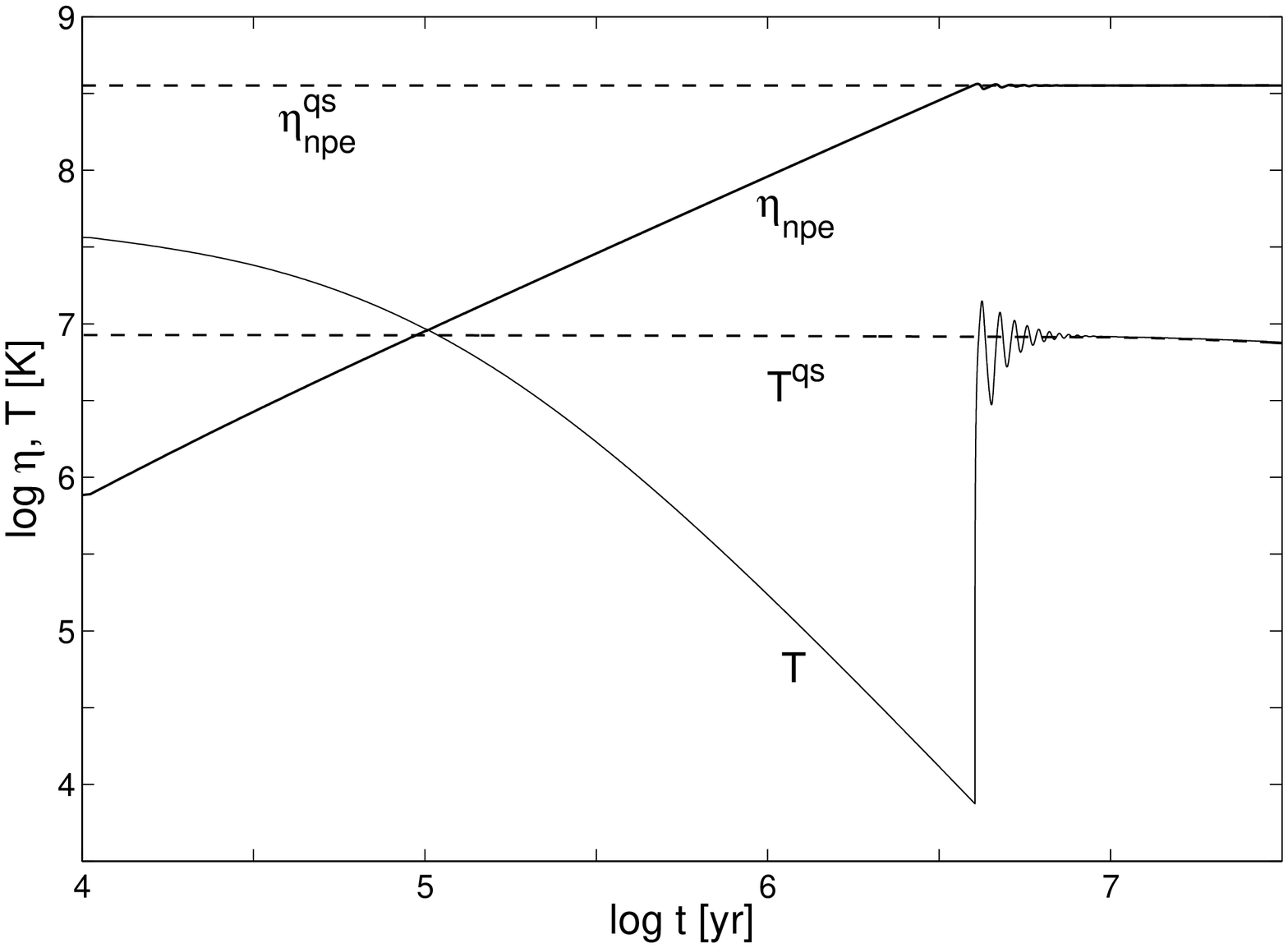}
\includegraphics[width=8.4cm,height=5cm]{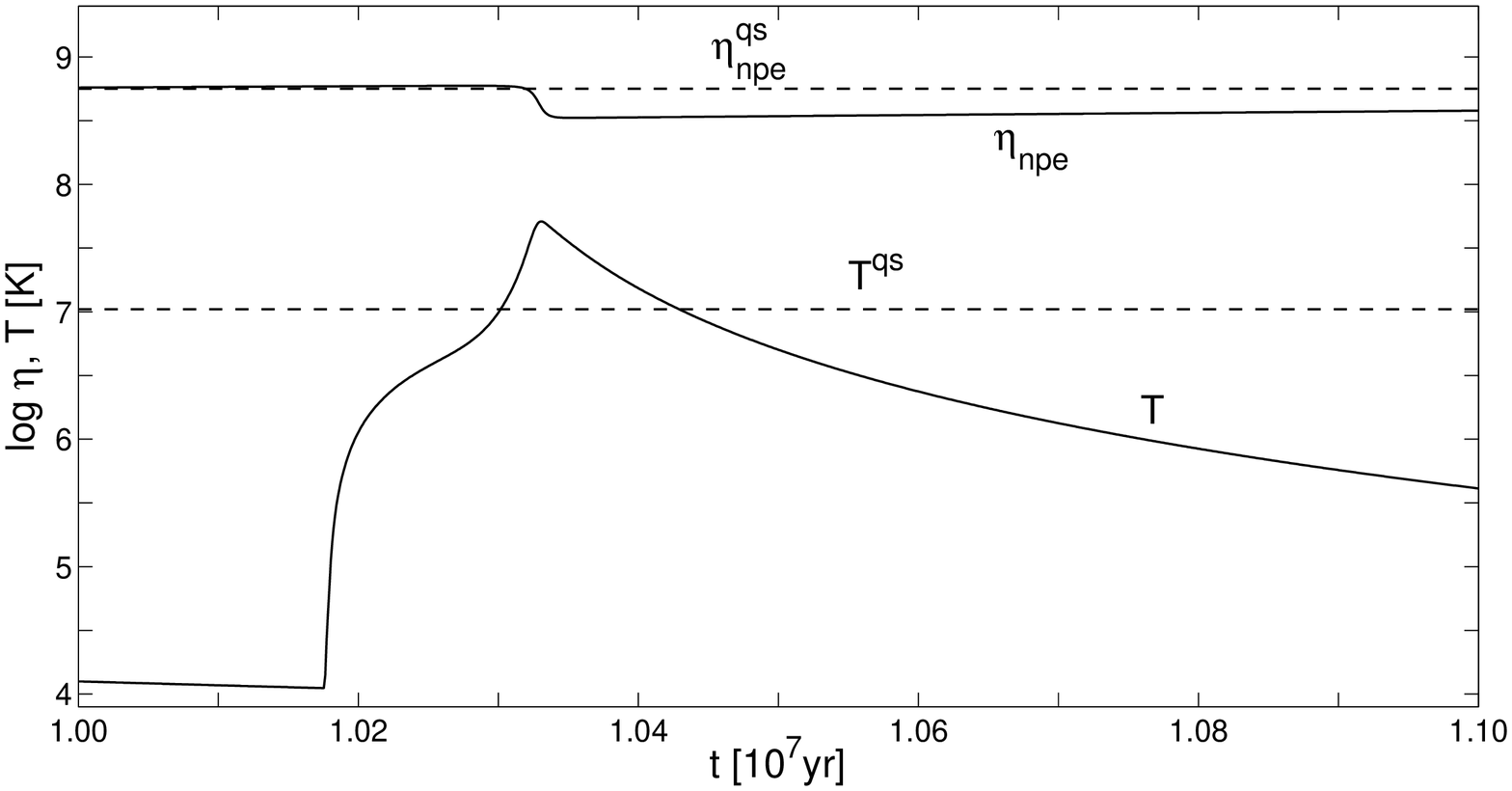} 
\end{tabular}
\label{fig:evolucion}
\caption{
Evolution of the internal temperature $T$ and the
chemical imbalance $\eta_{npe}$ for a 
1.68 $M_\odot$ star without muons built with the 
BPAL21 EOS \citep{pal88}, with the initial conditions $T=10^8$ K 
and $\eta_{npe}=0$. The spin-down is assumed to be caused by
magnetic dipole radiation, with dipolar field strength
 $B=2.8\cdot10^8$ G and initial period $P_0=1$ ms. The dashed lines
 are the quasi-steady state solution calculated from $\dot{T}=\dot{\eta}_{npe}=0$. 
In all cases, the protons are taken to be non-superfluid, while the neutron 
 superfluid energy gaps are  $\Delta_n=0.01$ MeV (\textit{upper left panel}),
$\Delta_n=0.03$ MeV (\textit{lower left panel}), and
$\Delta_n=0.05$ MeV (\textit{upper and lower right panels}). 
The \textit{lower right panel} shows a close-up of the second peak
in temperature of the evolution plot shown in the upper right panel 
($\Delta_n=0.05$ MeV)
with a linear scale in time.
}
\end{figure*}

In the upper right panel of Fig. 1, we use $\Delta_n=0.05$ MeV to show
strong oscillations without a clear damping effect. 
This behavior can be understood as follows.
Once the chemical imbalance increases to a critical value $\sim \Delta_{n}$,
the reaction rates increase dramatically, 
suddenly increasing the temperature and thus decreasing the imbalance. 
At this point, the imbalance is  substantially below the energy gap
and the direct Urca reactions are strongly inhibited (see \citealt{petro10} for more 
details on the reduction factors). 
Thus, the star begins to cool  almost exclusively
by photon emission, and the spin-down compression forces the chemical
imbalance to grow almost unimpeded by the neutrino reactions until
it reaches its critical value again. 
This process is repeated cyclically in a timescale given, essentially, by the time
taken by the chemical imbalance to come back to its critical value $\sim\Delta_{n}$,
where it erases any record of the previous cycles.

The lower right panel of Fig. 1, shows a close-up of one
of the peaks depicting the  evolution with $\Delta_{n}=0.05$ MeV 
in order to resolve and illustrate the rather sharp peaks. 
We also illustrate in this figure that  the timescale involved in the 
rapid increase in temperature
is on the order of $\sim 10^{5}$ yr, while the decrease happens on a timescale
of $\sim 10^{6}$ yr. 

This unexpected behavior will be the focus of our analysis.

\subsection{Stability analysis} \label{sec:results_stability}

In this section, we analyze the local stability of the temperature and
chemical imbalance by perturbing the evolution Eqs. 
(\ref{eq:evolucion_Ti}) and (\ref{eq:evolucion_eta1}) around the 
quasi-steady state  $\dot{T}=\dot{\eta}_{npe}=0$ and computing
the growth rates of these small displacements. 
For simplicity, we ignore the presence of muons in our analysis.
 
 We can safely neglect the modified Urca reactions of the electrons because
the direct Urca reactions are much more efficient and the superfluid suppression is
similar in both cases. Thus, the evolution equations (with  $\eta_{npe}$ 
in units of temperature) can be written as 
 \begin{eqnarray}
 \label{eq:evol_1}
	\dot{T} & =&\frac{1}{C}\left\{\tilde{L}_{D}\left[ \xi_{e} I_\Gamma
-I_\epsilon \right]T^6 - \tilde{L}_\gamma T^\alpha \right\} \equiv \mathcal{F}(T,\eta_{npe}),\\
 \label{eq:evol_2}
	\dot{\eta}_{npe} & = & -\frac{Z_{npe}}{k^2}\left(\tilde{L}_{D} I_\Gamma T^5\right) 
 + \frac{2W_{npe}}{k} \Omega \dot{\Omega} \equiv \mathcal{G}(T,\eta_{npe}). 
\end{eqnarray}

We considered the fully accreted envelope model of \citet{potetal97} to express 
the photon luminosity in terms of the internal temperature as 
$L_\gamma^\infty=\tilde{L}_\gamma T^\alpha$, where $\tilde{L}_\gamma$
is a numerical factor that depends on the stellar structure, and 
$\alpha=2.42$. Additionally,
we defined the functions $\mathcal{F}$ and $\mathcal{G}$ 
to save notation in the subsequent analysis.

We write the perturbed solutions to these equations as 
 \begin{eqnarray}
 \label{eq:perturbed_sol}
 T=T^{qs}+\delta T e^{\gamma t},\mbox{ and} \quad
\eta_{npe}=\eta_{npe}^{qs}+\delta \eta_{npe} e^{\gamma t},
\end{eqnarray}
where $T^{qs}$ and $\eta_{npe}^{qs}$ are the solutions of 
$\mathcal{F}=\mathcal{G}=0$, which are unique for each combination
of input parameters, i.e. energy gaps,
spin-down power, and structure constants.  We define $\gamma$ as the growth rate of the 
perturbations $\delta T$ and $\delta \eta_{npe}$. 
Replacing these perturbed solutions in the differential equations leads
to the usual eigenvalue problem for the growth rates of these perturbations
\begin{eqnarray}
\left(\gamma  \tens{I}- \tens{J}\right) 
\left( \begin{array}{cc} \delta T \\ \delta \eta_{npe} \end{array}   \right) =0,
\end{eqnarray}
 where $\tens{I}$ is the $2\times2$ identity matrix and $\tens{J}$ is the Jacobian matrix resulting 
 from the linearization of  Eqs. (\ref{eq:evol_1}) and (\ref{eq:evol_2}) and  evaluated at 
$T^{qs}$ and $\eta_{npe}^{qs}$. Thus, we write the Jacobian matrix elements,
 replacing the quasi-steady state and taking logarithmic derivatives, as 
 \begin{eqnarray}
 \label{eq:J_{11}}
\tens{J_{11}}&=&\frac{\partial\mathcal{F} } {\partial T} = \frac{\tilde{L}_\gamma}{\tilde{C}} T^{\alpha-1}  \left(\frac{\partial \ln (\xi_{e}I_\Gamma-I_\epsilon)}{\partial \ln T} +(6-\alpha)   \right),\\ 
\label{eq:J_{12}}
\tens{J_{12}}&=& \frac{\partial\mathcal{F} } {\partial \eta_{npe}}= \frac{\tilde{L}_\gamma}{\tilde{C}} T^{\alpha-1}  \left(\frac{\partial \ln (\xi_{e}I_\Gamma-I_\epsilon)}{\partial \ln \eta_{npe}} \right), \\
 \label{eq:J_{21}}
\tens{J_{21}} &=&\frac{\partial\mathcal{G} } {\partial T}=  \frac{2W_{npe}\Omega\dot{\Omega}}{ kT}
			 \left(\frac{\partial \ln (\xi_{e}I_\Gamma)}{\partial \ln T} +6  \right),\\
\label{eq:J_{22}}		
\tens{J_{22}}&=&\frac{\partial\mathcal{G} } {\partial \eta_{npe}} =\frac{2W_{npe}\Omega\dot{\Omega}}{ k\eta_{npe}} \left(\frac{\partial \ln (\xi_{e}I_\Gamma)}{\partial \ln \eta_{npe}} -1\right).
\end{eqnarray}

\begin{figure}[!h]
\includegraphics[width=8.5cm,height=6.3cm]{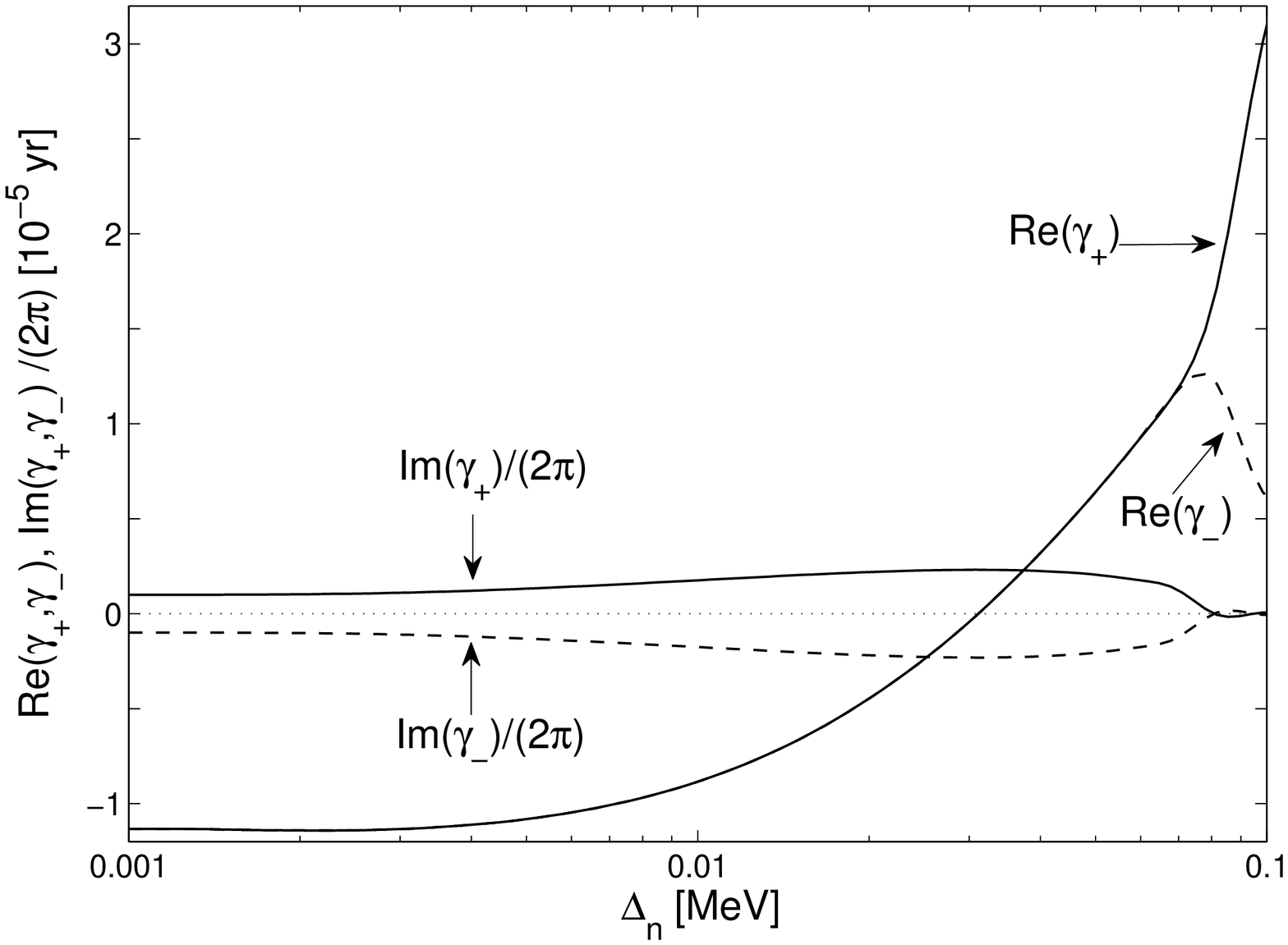}
\includegraphics[width=8.5cm,height=6.3cm]{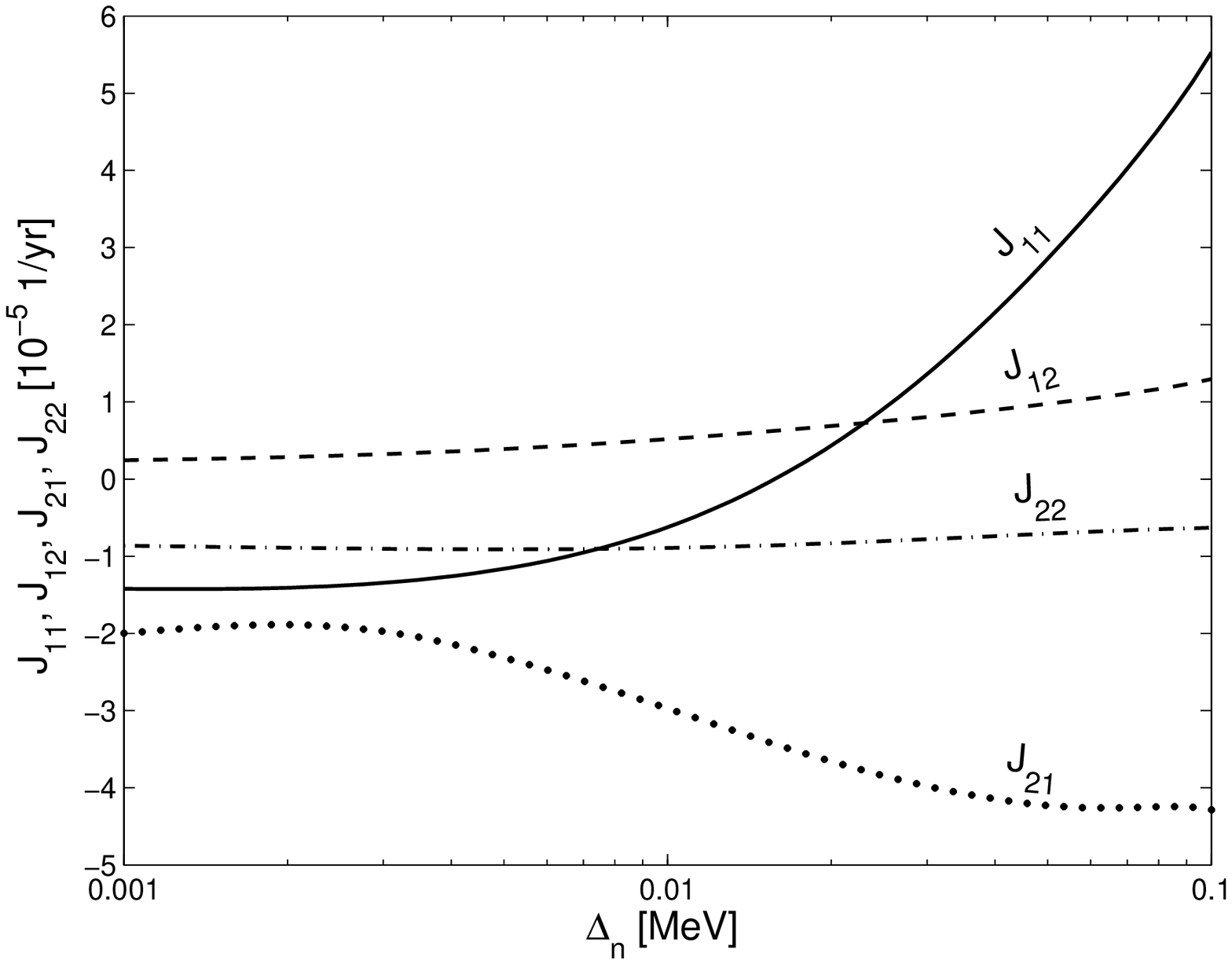}
\caption{Real and imaginary parts of the growth
rates $\gamma_{+}$ and $\gamma_{-}$ (upper panel) calculated from Eq. (\ref{eq:roots})
and components of the Jacobian matrix (lower panel) in Eqs. 
(\ref{eq:J_{11}}), (\ref{eq:J_{12}}), (\ref{eq:J_{21}}), and (\ref{eq:J_{22}}), 
for different values of $\Delta_{n}$. In both panels, the stellar parameters
are the same as in Fig. 1 and we assume that $\Omega\dot{\Omega}=-3\cdot10^{-9}[\mbox{rad}^2/\mbox{s}^3]$. }
\end{figure}
 
We solve for the two growth rates $\gamma_{+}$ and $\gamma_{-}$  by simply calculating
\begin{eqnarray}
\label{eq:roots}
\gamma_{\pm}&=&\frac{\tau\pm\sqrt{\tau^2-4\Delta}}{2}, 
\end{eqnarray}
where $\tau$ and $\Delta$ are the trace and the determinant of $\tens{J}$,
respectively. 
The stability of this system of differential equations, at least
when the linear approximations is valid, i.e. close to  quasi-steady solutions,
depends on these growth rates.
Thus, if $Re(\gamma_{\pm})<0$, the fixed point solutions will be stable, as happens
when we have only modified Urca reactions or no superfluid nucleons
are considered.
Otherwise, if $Re(\gamma_{\pm})>0$ the solutions will diverge. Moreover, the 
solution will oscillate if $Im(\gamma_{\pm})\neq0$, and the oscillations become important
during the evolution if $\left| Im(\gamma_{\pm}) \right|$ are comparable to or greater than
$\left| Re(\gamma_{\pm})\right|$.

In the upper panel of Fig. 2, we show the real and imaginary parts of the two growth rates 
for different values of
 $\Delta_n$. We fix the value of the spin-down parameters to
$\Omega\dot{\Omega}=-3\cdot10^{-9}[\mbox{rad}^2/\mbox{s}^3]$ in order
to compare our results to those of Fig. 1, where $\Omega\dot{\Omega}$
reaches this value at the age of $5\cdot10^{6}$ yr and slowly changes
until $\sim10^{8}$ yr.
When the energy gap is $\Delta_{n}=0.01$ MeV (top
panel of Fig. 1), the system
is locally stable because $Re(\gamma_{\pm})< 0$, and the oscillation is
highly damped because $|Re(\gamma_{\pm})| \gg |Im(\gamma_{\pm})|$.
Then, if we take $\Delta_{n}=0.03$ MeV (middle panel of Fig. 1),
we can see that the system is still locally stable, but 
$Re(\gamma_{+})$ and $Re(\gamma_{-})$ increase in such a way that
$|Re(\gamma_{\pm})| < |Im(\gamma_{\pm})|/(2\pi)$, and therefore the oscillations
are less damped than those with lower energy gaps.
Up to here, the linearized system predicts that the fixed point $T^{qs}$ 
and $\eta_{npe}^{qs}$  behaves as a \textit{stable spiral } (also known as an
attractor or a sink) in the $T-\eta_{npe}$ space,
hence the linear analysis ensures that it is also correct for 
the non-linear system close to this fixed point \citep{strogatz}.

Increasing the energy gap to a value higher than $\Delta_{n}=0.03$ MeV,
the linearized system becomes unstable because $Re(\gamma_{\pm})$ changes from 
being negative to positive.
Beyond this point, the non-linear system acts as an \textit{unstable spiral}
 (also known as a repeller or source) in the $T-\eta_{npe}$ space.
If the initial condition of the system of Eqs. (\ref{eq:evol_1}) and (\ref{eq:evol_2})
is close to the fixed point, it will diverge from it.  
Afterwards, the amplitude of the perturbations becomes large enough to make
 the non-linear terms significant, and the linear analysis is no longer valid.
We show in section Sect. \ref{sec:limit_cycle} that, for high energy gaps,
the highly non-linear system converges to a \textit{stable limit cycle}.

\subsection{Superfluidity driving the oscillations}
\label{sec:driving_osc}

The oscillations are exclusively caused by the appearance of the superfluid energy
gaps at the Fermi level of the nucleons.
This drastically changes the dependence of the neutrino emissivity and
the net reaction rate on the internal temperature and chemical imbalance,
which causes the oscillations, as we show in this section.

In the lower panel of Fig. 2, we show the previous effect by means of the components of the 
Jacobian matrix in Eqs. 
(\ref{eq:J_{11}}), (\ref{eq:J_{12}}), (\ref{eq:J_{21}}), and (\ref{eq:J_{22}}),
evaluated at the quasi-steady state.
From here, we can observe that close to the quasi-steady state,
 the evolution represented by Eqs. (\ref{eq:evol_1}) and (\ref{eq:evol_2}) 
 becomes unstable, essentially, because the component 
 $\tens{J}_{11}=\partial\mathcal{F}/\partial T$ changes from being negative
 to positive as we increase $\Delta_{n}$, changing the sign of the
 trace of the Jacobian $\tau$. 
 
 To understand why this happens, we analyze the dimensionless
 net heating function  $\xi_{e} I_\Gamma -I_\epsilon $ that appears
 in Eq. (\ref{eq:evol_1}). 
 For the non-superfluid case, one finds from the polynomials given in Eqs.
 (\ref{eq:F_D_def}) and  (\ref{eq:H_D_def}) that, in the quasi-steady state, where
 $T\ll \eta$,
 $\xi_{e} I_\Gamma -I_\epsilon \sim 21/(457\pi^6)\left(\eta/T\right)^{6}$, 
 which gives
$\partial \ln (\xi_{e}I_\Gamma-I_\epsilon)/\partial \ln T\sim-6$ , i.e. 
from Eq. (\ref{eq:J_{11}}) $\tens{J}$ remains negative. 
In contrast, for the superfluid case, we find that this function
 can increase with temperature for values of the energy gaps sufficiently 
 greater than the temperature. 
We know that both of the functions $I_\Gamma$ and $I_\epsilon$ increase 
with temperature since this increases the
phase-space available to the particles.
However, as \citet{petro10} showed by means of the reduction factors, 
superfluidity blocks the phase-space of the integrals $I_\epsilon$  and $I_\Gamma$ 
quite differently, and for the quasi-steady state we find that
$R_{\epsilon}\ll R_{\Gamma}$ (see Fig. 2 of \citet{petro10}), which forces
 $\xi_{e} I_\Gamma -I_\epsilon$ to be an increasing function of the temperature.
 Finally, this happens until the temperature is high enough to make the reduction 
factors comparable and the function decreases with temperature
in a similar way to its non-superfluid counterpart.

\subsection{Limit cycle in $\eta_{npe}-T$: existence and $\Delta$-dependence}
\label{sec:limit_cycle}

In the upper panel of Fig. 3, we show the evolution
of rotochemical heating in the $\eta_{npe}-T$ space
for different initial conditions
that converge to the same the limit cycle.
 Moreover, we know that the trajectories inside the limit cycle are 
repelled by the fixed point of $\mathcal{F}$ and $\mathcal{G}$, 
which in this case is an unstable spiral, as discussed in Sect. \ref{sec:results_stability}. 
Thus, we could  construct a trapping region, excluding the
fixed point,  such as the vector that field points inward at the 
boundary of this region and, therefore,  argue  by means of the Poincar\'e-Bendixson 
theorem that there is a closed orbit, i.e. a stable limit cycle 
(see, e.g. \citealt{strogatz} for details on limit cycles).

\begin{figure}[!h]
\includegraphics[width=8.5cm,height=6cm]{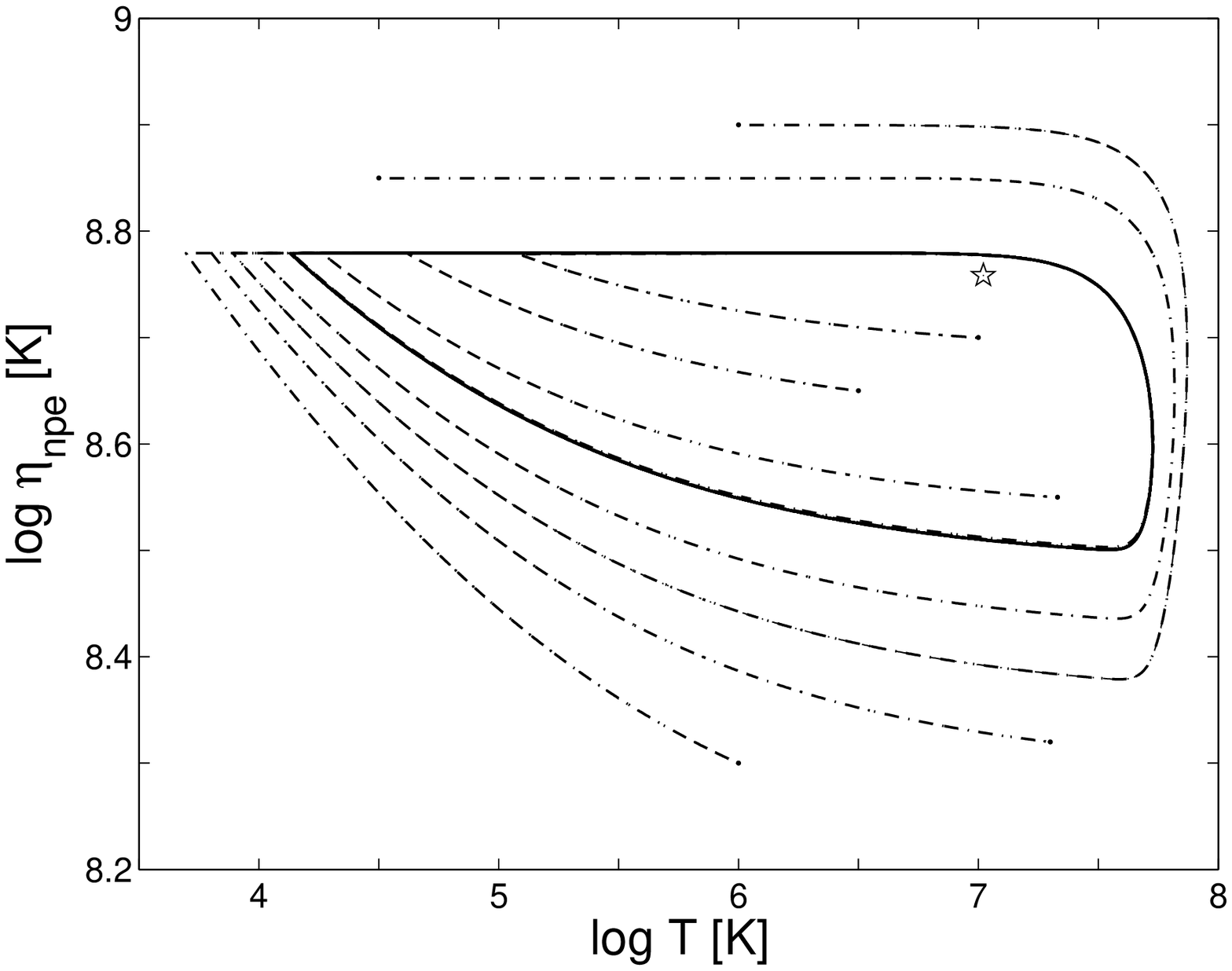}
\includegraphics[width=8.5cm,height=6cm]{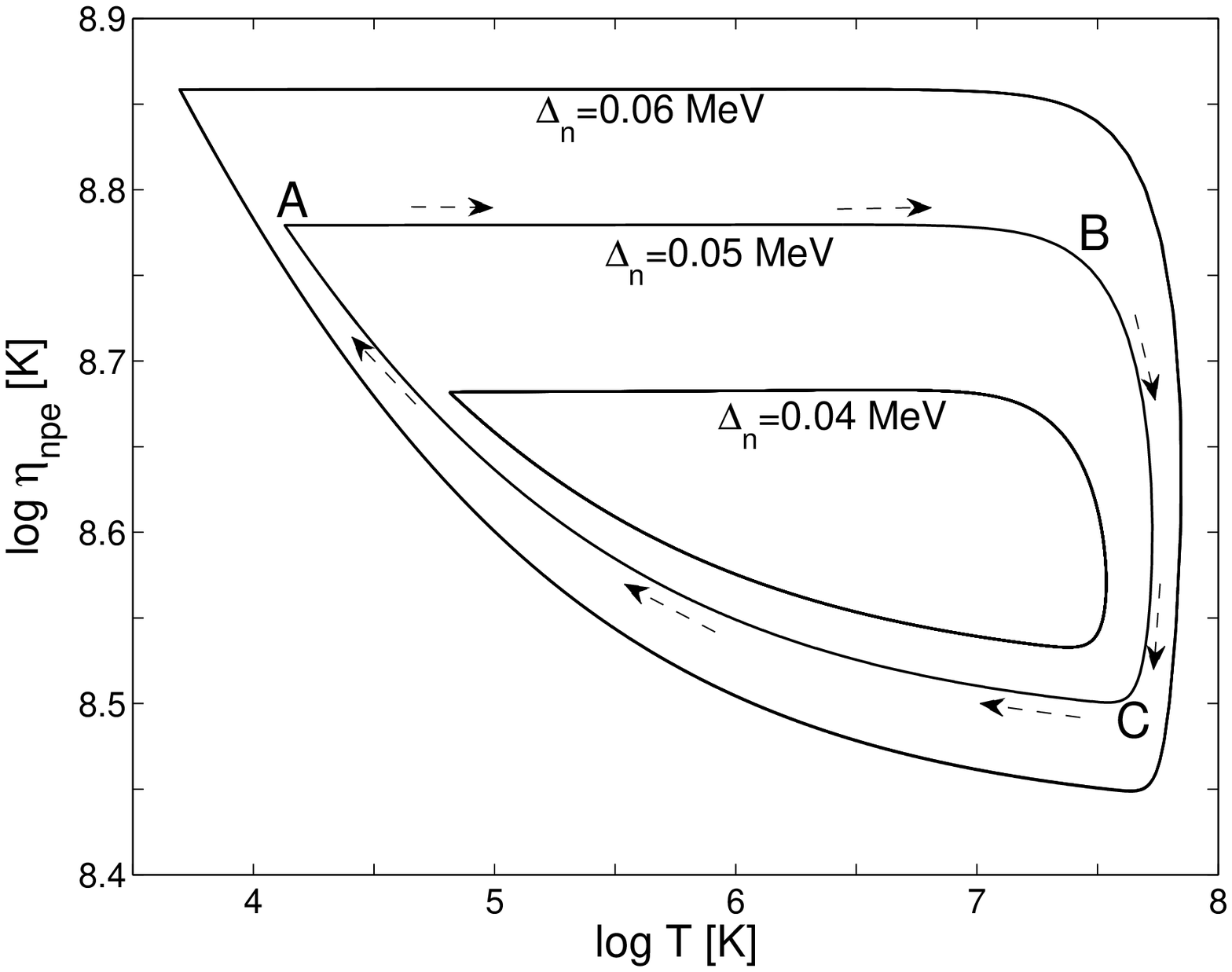}
\caption{Evolution of rotochemical heating in the $\eta_{npe}-T$ space for
a superfluid star with the same parameters of Fig. 1.
\textit{Upper panel:} evolution with  $\Delta_{n}=0.05$ MeV for different initial
conditions (dotted-dashed lines) that converge to the same limit cycle (solid line).
 The quasi-steady state is indicated (star).
\textit{Lower panel:}  evolution of the limit cycle for 
$\Delta_{n}=0.04$,  0.05,  0.06 MeV. 
The arrows indicate the path of the cycles A-B, B-C, and C-A for the
star with $\Delta_{n}=0.05$ MeV.
}
\end{figure}

In the lower panel of Fig. 3, we show different limit cycles 
by changing the energy gap  and label three sections of the evolution
 curve with $\Delta_{n}=0.05$ MeV, namely A-B, B-C, and C-A.
Along the path A-B, the chemical imbalance is $\sim\Delta_{n}$ and some reactions
are opened, increasing the temperature until it reaches node B, without
changing the chemical imbalance. Along the path B-C, 
the temperature is high enough to open abruptly many reactions
that tend to restore the beta equilibrium, and $\eta_{npe}$ drops rapidly.
Finally, along the path C-A, the chemical imbalance is sufficiently below the energy
gap to strongly inhibit the neutrino reactions. Thus, 
no heating mechanism is present and the star cools by photon
emission. Moreover, no restoring mechanism for the beta equilibrium
is present and the chemical imbalance grows by the spin-down 
compression.

We observe from this figure that the amplitude
of the limit cycles increases with the gap $\Delta_{n}$, which
we can easily understand from the previous analysis. 
For the path A-B, the chemical imbalance must be larger
 for larger $\Delta_{n}$ to open reactions.
Then, from  B to C,
we note from the linear analysis in Sect. \ref{sec:results_stability} 
 that for larger gaps the system becomes more unstable or, more precisely,
the heating function $\xi_{e} I_\Gamma -I_\epsilon$ becomes more 
sensitive to small perturbations in temperature reaching higher temperatures,
forcing the chemical imbalances to reach lower values in node C.
Finally, from C to A, all the curves are parallel since the photon
emission and spin-down compression do not depend on $\Delta_{n}$.

\subsection{Limit cycle timescales and the MSP J0437}\label{sec:timescale}

We can observe from the bottom panel of
Fig. 1 that there are two very different scales
governing the limit cycles. 
The first and shorter one involves the rapid increase in temperature
and ends with a rapid decrease in the chemical imbalance, i.e. it
goes from A to C in the lower part of Fig. 3. 
The second scale goes from C to A and is substantially dominant
in the rotochemical heating evolution.

The first  is quite stable and on the order of the timescales shown
in Fig. 2 for the growth rates $\gamma$, i.e. a timescale of $\sim10^{5}$ yr.
However, the second scale depends strongly on the energy gap, because the latter
sets the amplitude of the drop in $\eta_{npe}$. It also depends on 
$\Omega\dot{\Omega}$,  since it accounts for the rate at which the 
chemical imbalance grows.
For the examples displayed in Figs. 1 and 3, the longer timescale
is  $\sim10^{6}-10^{7}$ yr.

Considering the large difference between both timescales, we check that
the fraction of the time in which the thermal emission would be detectable, i.e. the
surface temperature $T_s\gtrsim 10^5$ K, is in the range
$\sim5-25\%$, mainly depending on the energy gaps and spin-down
parameters. 
Moreover, it is also unlikely to find it close to its quasi-steady state
and therefore are unable to draw further conclusions about 
the pairing gaps needed to explain observations as in \citet{petro10}.

However, we can account for the maximum energy gap combination
$\Delta_{thr}=\Delta_{n}+\Delta_{p}$ at which rotochemical heating
evolves to a stable solution, i.e. its quasi-steady state, and
predict a surface temperature.
Thus, for the spin-down parameters of MSP J0437 and the set of EOSs 
from \citet{pal88} that allow direct Urca reactions of electrons and 
muons for its mass range  $1.76\pm0.2$ M$_\odot$  \citep{deller},
namely BPAL21, BPAL31, BPAL 32, and BPAL33, the maximum
gap combination that predict a quasi-steady state
is in the range of $\Delta_{thr}\sim0.01-0.1$ MeV.
Thus, the temperatures predicted when including superfluidity
are a few times higher than those predicted by the non-superfluid
case \citep{FR05}.
They are quite close to $10^{5}$ K and
for some cases can explain the likely thermal emission
of the MSP J0437 \citep{kargaltsev04}.

Finally, given the rapid increase in temperature,
 we could ask whether the finite thermal diffusion timescales are 
relevant to our study or  make any difference to the validity of our assumption
of an isothermal NS interior. We claim that this assumption is
fairly safe, since the increase in temperature takes
$\sim10^{5}$ yr, while the thermal diffusion time is just a few decades
for these low temperatures.

\begin{figure}[!h]
\includegraphics[width=8.5cm,height=6.3cm]{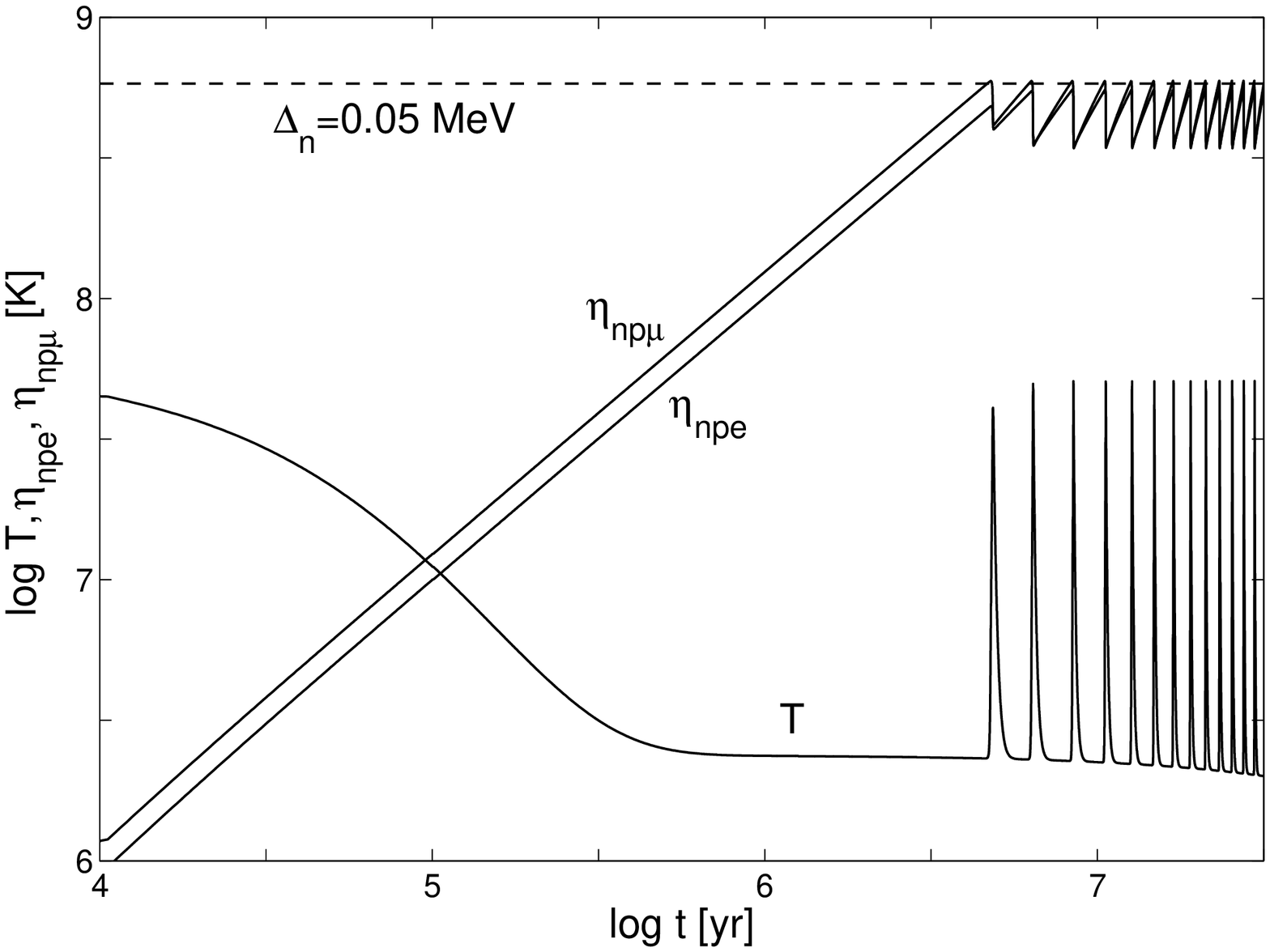}
\includegraphics[width=8.5cm,height=6.3cm]{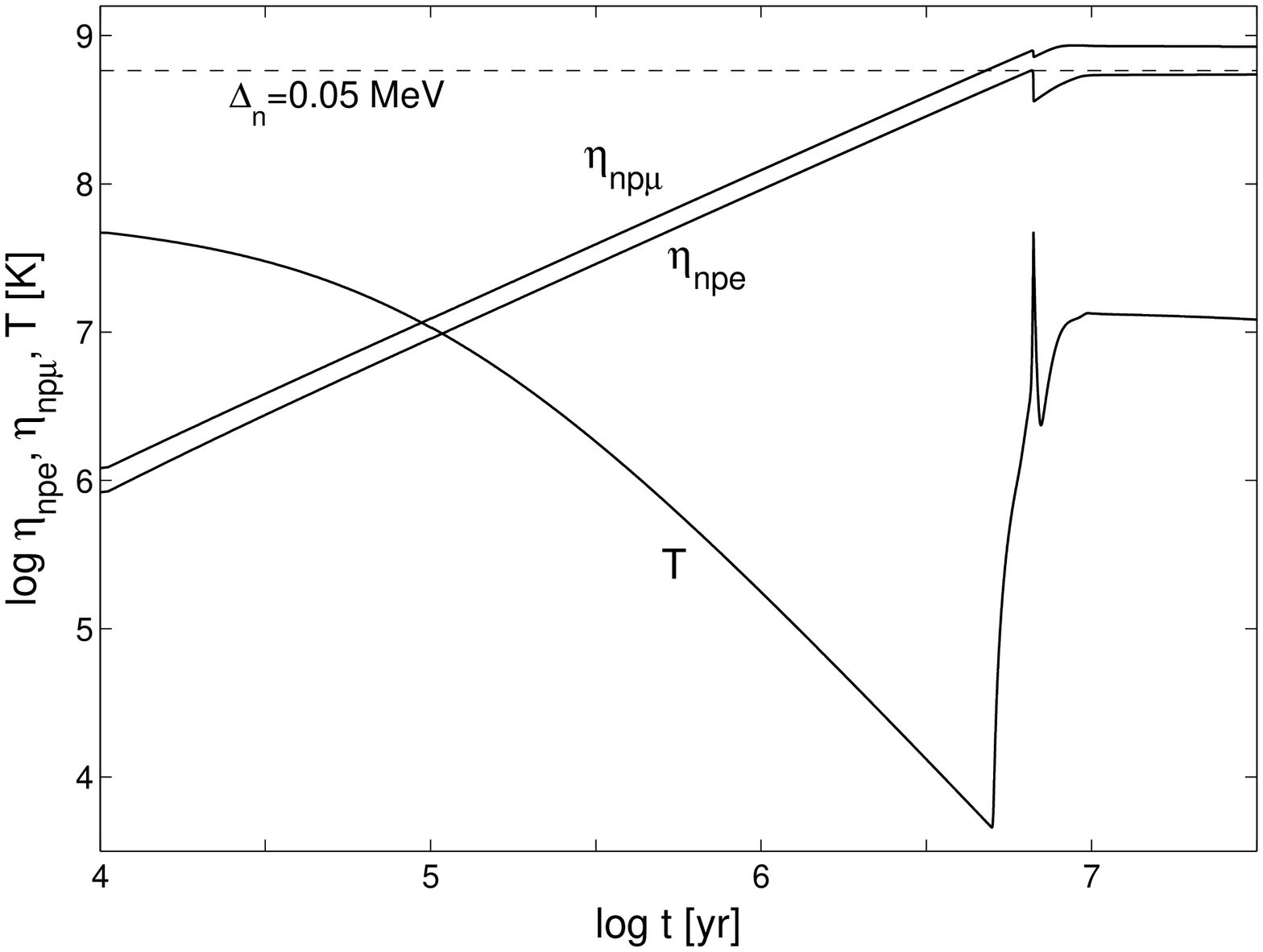}
\caption{
Evolution of the internal temperature $T$ and the
chemical imbalances $\eta_{npe}$ and $\eta_{np\mu}$
 with the initial condition $T=10^8$ K 
and null chemical imbalances. The spin-down is assumed to be caused by the 
magnetic dipole radiation, with dipolar field strength
 $B=2.8\cdot10^8$ G and initial period of $P_0=1$ ms. The dashed lines
indicate the energy gap $\Delta_{n}=0.05$ MeV.
\textit{Upper panel:} evolution of a 1.69 $M_\odot$ star built with the 
BPAL21 EOS \citep{pal88}, which allows muon direct Urca reactions, and we
add the vortex creep reheating term of  Eq. (\ref{eq:J}) with
an excess of angular momentum $J=10^{43}$ erg s.
\textit{Lower panel:} evolution of a 1.68 $M_\odot$ star built with the 
BPAL21 EOS \citep{pal88}, which forbids direct Urca reactions with muons.
}
\end{figure}

\subsection{Effect of muons and other reheating mechanisms.}

The presence of muons does not introduce changes 
to the previous analysis, as long as direct Urca reactions
with muons are allowed. On the other hand, if they react
only by means of the modified Urca channels, the oscillations are damped and
finally vanish.

In the upper panel of Fig. 4, we observe that the only effect to be considered is that
the oscillations would be driven by the reactions involving
 muons, instead of those with electrons. 
This happens because the departure from beta equilibrium with
muons, i.e. $\eta_{np\mu}$, is more strongly affected by the
spin-down compression, and the direct Urca reactions with
muons will lead the oscillations since they reach the critical
value $\sim\Delta_{n}$ first.

Moreover, including another reheating mechanism does not
alter the existence of the oscillations as long as the temperature
increase produced by this mechanism is not so high as to
overcome the quasi-steady value of the temperature.
As \citet{denis10} argued, the two most relevant mechanisms to reheat an
old neutron star are rotochemical heating and vortex creep (\citealt{alpar},
\citealt{shibazaki}).
Therefore we include the latter in the upper panel of Fig. 4 to show that the
oscillations persist even when another reheating mechanism is included.
The energy dissipation rate that must be included as a heating term in Eq. 
(\ref{eq:evolucion_Ti}) is given by \citep{alpar}
\begin{equation}
\label{eq:J}
L_{vortex}=J\dot{\Omega},
\end{equation}
where $J$ is the excess of angular momentum.
By choosing an intermediate value of $J=10^{43}$ erg s for the models
in \citet{shibazaki}, we observe that
the  only effect of the extra heating term is to provide a higher floor to the internal temperature. 
 This results in a straightforward way from
the analysis in Sect. \ref{sec:results_stability}, since the Jacobian element in
Eq. (\ref{eq:J_{11}}) is unaltered by the new term in Eq. (\ref{eq:J}).
However, if this mechanism keeps the NS at substantially higher
 temperatures, it could displace the quasi-steady state and change the behavior
of the oscillations.

In the lower panel of Fig. 4, we show the evolution when
direct Urca reactions with muons are forbidden, but modified Urca reactions
involving muons are present.
At first,  the chemical imbalance of the muons reaches a value 
close to the energy gaps and reheats the star by means of
modified Urca reactions. 
The chemical imbalance of the electrons then reaches this threshold
value and the oscillations start to work normally until
the chemical imbalance of the muons reaches its quasi-steady state, 
which predicts higher temperatures  than those without 
muons (see the bottom panel of Fig. 1). 
Hence, the oscillations vanish and the quasi-steady state is 
mostly governed by the chemical imbalance of the muons.

\begin{figure}[!h]
\includegraphics[width=8.5cm,height=6.3cm]{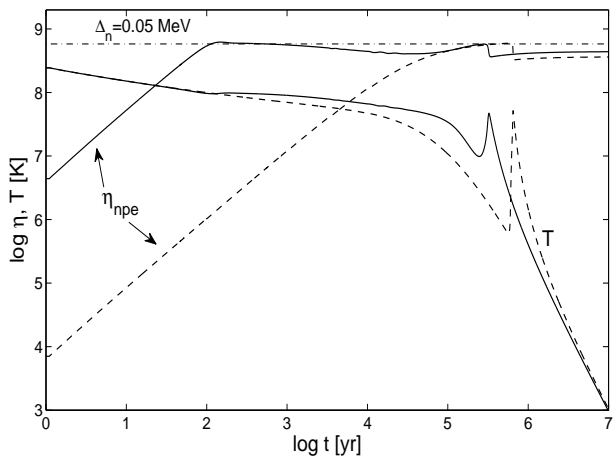}
\caption{
Evolution of the internal temperature $T$ and the
chemical imbalance $\eta_{npe}$ for 
a 1.68 $M_\odot$ star built with the EOS BPAL21 and no muons,
 with the initial condition $T=10^{9}$ K 
and null chemical imbalance. 
The spin-down is assumed to be caused by the 
magnetic dipole radiation, with dipolar field strengths
$B=10^{11}$ G (dashed line) and $B=10^{12}$ G (solid line) 
 initial period of $P_0=5$ ms. The dotted-dashed line
indicates the energy gap $\Delta_{n}=0.05$ MeV.}
\end{figure}

\subsection{Effect on classical pulsars}\label{sec:classical}

In this section, we show the effects of these oscillations
on classical pulsars.
In Fig. 5, we illustrate their unstable behavior by showing
the evolution of rotochemical heating for a NS with an energy gap 
$\Delta_{n}=0.05$ MeV, electron direct Urca reactions, no muons, and two 
magnetic fields of  $B=10^{11}$ G and $B=10^{12}$ G.

From this figure, we observe that the temperature and the chemical
imbalance can oscillate only during half of the period
for $B=10^{11}$ G or a full period for $B=10^{12}$ G
since the rotation rate is insufficient for the imbalance to recover 
from its drop and, therefore, produce any other oscillation.
Thus, after the sudden increase in temperature at $\sim10^{5}$ yr, 
the star cools almost exclusively by means of photon emission, while the
chemical imbalance grows slowly, never reaching the threshold 
of the energy gap again.

If we changed the initial period to lower values (e.g. $P_{0}=1$ ms),
we could observe from the case with  $B=10^{11}$ G 
one full oscillation period and another half, similar to the lower panel of
Fig. 4. Similarly, lower values of the energy gap could produce more 
oscillations.
In contrast, if we increase the initial period, the rotation energy 
to increase the chemical imbalance
may not be enough to reach the energy gap threshold, and oscillation
may not occur at all.

In conclusion, the oscillating behavior generally does not persist in classical pulsars,
and the standard cooling governs their thermal evolution.

\section{Conclusions} \label{sec:conclusions}

We have studied  a new kind of oscillations of the 
temperature and chemical imbalances, produced during the 
evolution of rotochemical heating when the direct Urca reactions 
are allowed in superfluid MSPs with relatively high uniform and 
isotropic Cooper pairing gaps.

The oscillations work as follows.
The direct Urca reactions are strongly
blocked until the chemical imbalances grow up to a threshold value 
$\Delta_{thr}=\Delta_{n}+\Delta_{p}$.
Soon after this happens, strong reactions are suddenly turned on, increasing
the temperature and decreasing the chemical imbalances.
As this happens, the reactions are again strongly blocked and the star
cools down by photon emission, while the chemical imbalances
again increase until they reach the threshold value required to repeat the cycle
after $10^{6-7}$ yr. The temperature stays high for only a small fraction
of the cycle, making it difficult to detect this effect and predict a NS
temperatures at a given time. For gaps below $\sim0.05$  MeV or with muons
reacting only by modified Urca processes, the oscillations vanish
and the system reaches a quasi-steady state, as previously found
in the non-superfluid or modified Urca cases.

\begin{acknowledgements}
We thank  Denis Gonz\'alez and Nicol\'as
Gonz\'alez for discussions and 
comments that benefited the present paper, and 
Rodrigo Fern\'andez for letting us use his  
rotochemical heating code.
This work was supported by Proyecto Regular FONDECYT 
1060644, the FONDAP Center of Astrophysics (15010003),
 Proyecto Basal PFB-06/2007, 
and Proyecto L\'imite VRAID No 15/2010.
\end{acknowledgements}

\end{document}